\documentstyle[preprint,aps]{revtex}
\draft
\begin{document}
\title{Efficient Peltier refrigeration by a pair of normal metal/
insulator/superconductor junctions}
\author{M.M. Leivo and J.P. Pekola}
\address{Department of Physics, University of Jyv\"{a}skyl\"{a},
P.O. Box 35, 40351 Jyv\"{a}skyl\"{a}, Finland }
\author{D.V. Averin }
\address{Department of Physics, SUNY Stony Brook, NY 11794}
\maketitle

\begin{abstract}
We suggest and demonstrate in experiment that two normal
metal /insulator/ superconductor (NIS) tunnel junctions combined
in series to form a symmetric SINIS structure can operate as an
efficient Peltier refrigerator. Specifically, it is shown that
the SINIS structure with normal-state junction resistances
1.0 and 1.1 k$\Omega$ is capable of reaching a temperature of
about 100 mK starting from 300 mK. We estimate the corresponding
cooling power to be 1.5 pW per total junction area of
0.8 $\mu$m$^2$ at $T= 300$ mK.
\end{abstract}

\narrowtext
\newpage

Recently it was shown \cite{b1} that the Peltier effect in
normal metal /insulator/ superconductor (NIS) junctions can be used
to cool electrons in the normal electrode of the junction below
lattice temperature. The cooling arises due to the energy gap in the
superconductor, because of which quasiparticles with higher energy
are removed more effectively from the normal metal than
quasiparticles with lower energy. The decrease in electron
temperature demonstrated in this first experiment was, however,
limited to about 10\% of the starting temperature.
There are two possible reasons for this. The first and most
obvious, is that the resistance of the refrigerator junction was
relatively large leading to low cooling power. The second possible
reason is heat leakage through the SN contact used to bias
the refrigerator. Nominally, an ideal SN contact with large
electron transparency should be able to provide electric conductance
without thermal conductance at low temperatures. However both finite
temperature and finite subgap density of states in a superconductor
lead to non-zero thermal conductance of the biasing contact which
degrades the refrigerator performance. This poses the problem of how
to bias the refrigerator without compromising its thermal insulation.

In what follows we address these two problems and show that once they
are solved the refrigerator performance is improved dramatically. In
particular, we show that such a refrigerator is capable of reaching
temperatures of about 100 mK starting from 300 mK. This brings the
NIS refrigerators quite close to practical applications, for instance,
in cooling space-based infrared detectors \cite{b2}.

The problem of refrigerator junction resistance can be alleviated to
some degree in a straightforward way either by increasing the junction
area or decreasing the specific resistance of the insulator barrier.
Because  of the limitations of the fabrication method (see below)
the junction area could not be increased by much, so that the only
option available to us was to make the insulator barrier
thinner. Although it is a challenging technological problem to push
this process to its limit, we could conveniently reduce the specific
resistance of the junctions to about 0.3 k$\Omega\, \times\,
\mu$m$^2$.

It is less obvious how to solve the second problem of heat
leakage through the biasing junction. The solution
we suggest here is to combine two NIS junctions in series to form a
symmetric SINIS structure. Since the heat current in the NIS
junction is a symmetric function of the bias voltage $V$, the heat
flows out of the normal electrode regardless of the direction of
the electric current if the junction is biased near the tunneling
threshold, $V\simeq \pm \Delta/e$. This means that in a symmetric
SINIS structure we can realize the conditions when the electric
current flows into the normal electrode through one junction and
out through the other one, while the heat flows out of the normal
electrode through both junctions. In this way the heat leakage into
the normal electrode of the structure is minimized. The experiment
with the SINIS structures described below supports this idea.

We begin by briefly outlining the basic theoretical concepts
concerning the heat flow in the NIS junctions. Under typical
conditions when the transparency of the insulator barrier is
small, the heat current $P$ out of the normal electrode (cooling
power) of an individual NIS junction is:
\begin{equation}
P(V) =\frac{1}{e^{2} R_{T}} \int_{-\infty}^{+\infty} d\epsilon
N(\epsilon) (\epsilon -eV) [f_1(\epsilon-eV)-f_2(\epsilon) ]  \, ,
\label{1} \end{equation}
where $R_T$ is the normal-state tunneling resistance of the barrier,
$f_j$ is an equilibrium distribution of electrons in the $j$th
electrode, and $N(\epsilon) = \Theta (\epsilon^2 -\Delta^2) \mid \!
\epsilon \! \mid/\sqrt{\epsilon^{2} -\Delta^{2}}$ is the density of
states in the superconductor. From eq.\ (\ref{1}) we can deduce
several properties of the cooling power $P$. First of all, by
changing the integration variable $\epsilon \rightarrow -\epsilon$
we prove that for equal temperatures of the two electrodes $P$ is
indeed a symmetric function of the bias voltage,
$P(-V)=P(V)$. Plotting eq.\ (\ref{1}) numerically one can see that
$P$ is maximum at the optimal bias points $V\simeq \pm \Delta/e$.
The optimal value of $P$ depends on temperature and is maximum at
$k_B T\simeq 0.3\Delta$, when it reaches $0.06\Delta^2/e^2R_T$, and
decreases at lower temperatures as $(k_B T/\Delta)^{3/2}$ \cite{b3}.
(The last conclusion is different from the $T^2$ law obtained
in Ref.\ \cite{b1}.)
Specifically, at $V=\Delta/e$ (i.e., quite close to the optimal
bias voltage) one can get from eq.\ (\ref{1}):
\begin{equation}
P(\Delta/e) =\frac{\sqrt{\pi} (\sqrt{2}-1)}{4} \zeta(3/2)
\frac{\Delta^2}{e^{2} R_{T}} (\frac{k_B T}{\Delta})^{3/2} \simeq
0.48 \, \frac{\Delta^2}{e^{2} R_{T}} (\frac{k_B T}{\Delta})^{3/2} \, .
\label{2} \end{equation}

Figures 1a and 1b show, respectively, a schematic diagram of the
SINIS structures studied in our experiments and the corresponding
AFM image of the structure. Four tunnel junctions were fabricated
around a normal metal (Cu) central electrode and four
superconducting (Al) external electrodes. The electrodes were made
with electron beam lithography using the shadow mask evaporation
technique. The tunnel junctions were formed by oxidation in pure
oxygen between the two metallization steps. Two junctions at the
edges with larger areas were used for refrigeration, while the pair
of smaller junctions in the middle
was used as a thermometer. A floating measurement of voltage across
the two thermometer junctions at a constant bias current was used to
measure temperature in the same way as in the simpler one junction
case \cite{b4}.

Prior to our experiments with the SINIS refrigerator we repeated
measurements in the geometry of Nahum et al. \cite{b1}. In our
case the refrigerating junction had a resistance of $R_T =$ 7.8
k$\Omega$, the copper island was 10 $\mu$m long, 0.3 $\mu$m wide
and 35 nm thick. Results of the measurements of the electron
temperature in the island as a function of refrigerator voltage
$V_{refr}$ for several starting temperatures at $V_{refr}=0$ are
shown in Fig. 2. We see that only a few per cent refrigeration
can be obtained, as expected.

Solid lines in Fig.\ 2 represent a theoretical fit obtained within
the standard model of electron energy relaxation \cite{b6,b7,b8}.
Within this model, we assume that the electron-electron collision
rate is large so that the electrons maintain an equilibrium
distribution characterized by the temperature $T$, which is in
general different from the lattice temperature $T_l$. The rate of
energy transfer from electrons to phonons is then \cite{b7}:
$P_l=\Sigma U (T^5-T_l^5)$, where $\Sigma$ is a constant which
depends on the strength of electron-phonon coupling, and $U$ is the
volume of the island. Another element of the fitting process is
a heat conductance $\kappa$ of the biasing SN contact. (For
simplicity, we neglect temperature dependence of $\kappa$, since
the temperature range of interest in Fig.\ 2 is not very large.)
The value of the superconductor gap $\Delta$ is almost fixed
by the position of the temperature dips in the refrigeration
curves (Fig.\ 2) and is taken to be
155 $\mu$eV for this sample. Solving numerically the equation
$P=P_l+\kappa (T-T_l)$, where $P$ is the cooling power (\ref{1})
we can calculate $T$ as a function of $V_{refr}$. The fit in
Fig.\ 2 is obtained in this way with $\Sigma=0.9$ nW/K$^5\,
\mu$m$^3$ and $\kappa= 8$ pW/K. For comparison, the dashed line
shows the fit obtained for the lowest-temperature curve without
$\kappa$; in this case $\Sigma =1.4$  nW/K$^5\, \mu$m$^3$. Although
there is no drastic disagreement with the data even in this case,
we see that $\kappa$ improves the fit considerably.

To see whether the value of the heat conductance $\kappa$ deduced
from the fit in Fig.\ 2 is reasonable, we calculated the heat
conductance of the SN contact with perfect electron transparency
at low temperatures, using the method developed in \cite{b8,b3}:
\begin{equation}
\kappa = \frac{2\Delta}{e^2R_N} (\frac{2\pi\Delta}{k_B T})^{1/2}
\exp (-\frac{\Delta}{k_B T}) \, .
\label{3} \end{equation}
Making use of the gap value and temperature corresponding to Fig.\
2 we get from this equation that $\kappa=8$ pW/K corresponds to
the normal-state contact resistance $R_N$ on the order of 100 Ohm,
which is of the same order of magnitude as in the experiment.

Figure 3 shows our main results with the SINIS refrigerator of
Fig.\ 1. The two refrigerating junctions had resistances $R_T=$
1.0 and 1.1 k$\Omega$, respectively, and the island was 5 $\mu$m
long, 0.3 $\mu$m wide, and 35 nm thick. In Fig.\ 3  we see several
refrigeration curves starting at various ambient temperatures at
$V_{refr} =0$. Note that maximum cooling power is now obtained at
$V_{refr} \simeq 2 \Delta/e$ because of two junctions involved. From
the position of the temperature dips we get $\Delta = 180\; \mu$eV.
We see that the drop in temperature is immensely improved over
that of the single junction configuration. Solid lines in Fig.\ 3
show the theoretical fit which was obtained within the same model as
for the single-junction
configuration with the two modifications. We do not have heat
conductance $\kappa$ this time, and we need to solve the balance
equations simultaneously for the energy and electric current in
order to determine the electric potential of the island. The best
fit shown in Fig.\ 3 corresponds to $\Sigma = 4$  nW/K$^5\,
\mu$m$^3$. The fit can be classified as reasonable, although there
are some obvious discrepancies between the data and the theory.
Possible origins of these discrepancies include an oversimplified
model of electron-phonon heat transfer on the theory side, and poor
calibration of the thermometer toward higher temperatures on the
experimental side. The inset in Fig.\ 3 shows the maximum cooling
power as a function of temperature deduced from this fit, together
with the analytical dependence obtained by summing eq.\ (\ref{2})
over the two junctions. We see that the simple analytical expression
(\ref{2}) gives a very accurate description of the lower-temperature
cooling power.

We note in passing that the experiment described above provides a
more transparent alternative interpretation of the previous
experiments on the enhancement of superconductivity in the SIS'IS
structures \cite{b9}. It implies that the reason for the stimulation
of superconductivity in the central electrode of the SIS'IS
structure might not be the formation of the non-equilibrium
distribution of quasiparticles inside this electrode
\cite{b9,b10}, but simply a decrease in its temperature.

Before concluding we would like to mention that the next
step in the development of a practical NIS refrigerator could be
further optimization of the refrigerator performance with respect
to the resistance $R_T$ of the insulator barrier. As we saw above,
the cooling power of the refrigerator increases with decreasing
$R_T$. For a fixed junction area this trend would continue only
up to an optimal resistance, at which point the transport starts to
be dominated by Andreev reflection \cite{b3}. The theoretical limit
for the maximum cooling power density is on the order of $10^{-8}$
$W/\mu$m$^2$ for aluminum junctions and should be reached in the
junctions with unrealistically low specific resistances on the
order of $10^{-2}$ $\Omega\, \times \mu$m$^2$. In practice the
limiting
factors will be the technological ability to fabricate uniform
tunnel barriers with high transparency, and, conceivably, the
heat leakage through these tunnel barriers to the central island.

In conclusion, we have shown that the nominally-symmetric SINIS
structure can be used as an efficient Peltier refrigerator. One of
the advantages of the symmetric structure is that it is easier to
fabricate than the asymmetric single-junction configuration. Besides
this, SINIS structure provides more efficient thermal insulation
of the central electrode, which allowed us to demonstrate a
temperature drop of about 200 mK starting from 300 mK. The achieved
cooling power density was approximately 2 pW/$\mu$m$^2$, with the
total power being 1.5 pW at $T= 300$ mK.

\vspace{1ex}

The authors gratefully acknowledge useful discussions with K.
Likharev and M. Paalanen. This work was supported in part by the
Academy of Finland and ONR grant \# N00014-95-1-0762.

\figure{ Figure 1.} (a) The schematics of the SINIS refrigerator
used in the measurements, and (b) an AFM image of the actual
structure.

\figure{ Figure 2.} Results of the NIS single junction experiment:
temperature $T$ [mK] of the N-electrode versus refrigerator
voltage $V_{refr}$ [$\mu$V]. Dots and solid lines show, respectively,
experimental data and the theoretical fit including heat conductance
of the biasing contact. For comparison, the dashed line shows the fit
without the heat conductance.

\figure{ Figure 3.} SINIS refrigerator performance at various
starting temperatures. Notations are the same as in Fig.\ 2.
Dots are the experimental
data, while the solid lines show the theoretical fit with
one fitting parameter $\Sigma$ for all curves. The inset
shows the cooling power $P$ [pW] from the fits (dots), together
with the analytical result from eq.\ \protect (\ref{2}).

\end{document}